\documentclass{article}
\usepackage{amsfonts}
\usepackage{amsmath}
\usepackage{amssymb}
\usepackage{graphicx}%
\providecommand{\U}[1]{\protect\rule{.1in}{.1in}}

\begin{document}

\title{New Class of Generalized Extensive Entropies for Studying Dynamical Systems
in Highly Anisotropic Phase Space.}

\author{Giorgio SONNINO$^{1,2}$\\ $^1$Universit{\'e} Libre de Bruxelles (ULB), Department of Physics \\ Bvd du Triomphe, Campus de la Plaine CP 231\\ 1050 Brussels, Belgium. Email : gsonnino@ulb.ac.be\\
$^2$Royal Military School (RMS), Av. de la Renaissance 30, \\ 1000 Brussels, Belgium\\
and\\
\and Gy\"{o}rgy STEINBRECHER$^3$\\$^3$Association EURATOM-MEdC, University of Craiova,\\A. I. Cuza 13, 200585 Craiova, Romania \\Email: gyorgy.steinbrecher@gmail.com }

\maketitle

\begin{abstract}
Starting from the geometrical interpretation of the R\'{e}nyi entropy, we
introduce further extensive generalizations and study their properties. In
particular, we found the probability distribution function obtained by the
MaxEnt principle with generalized entropies. We prove that for a large class
of dynamical systems subject to random perturbations, including particle
transport in random media, these entropies play the role of Liapunov
functionals. Some physical examples, which can be treated by the generalized
R{\'e}nyi entropies are also illustrated.

\end{abstract}

Generally, to characterize some unknown events with a statistical model, it is chosen the one that has maximum entropy (MaxEnt principle) {\it i.e.}, the one that has maximum uncertainty. The corresponding probability distribution functions (PDF) are obtained by maximizing entropy, under a set of constraints. When the system is close to the thermodynamic equilibrium, the PDF may be obtained by extremizing the entropy production subject to restrictions. For this reason, the entropy production for plasmas with electromagnetic turbulence was analyzed and carefully calculated in many paper, by using the standard definition of the Boltzmann-Gibbs theory \cite{garbet}-\cite{sugama}. In other terms, entropy is regarded as a measure of information, and it is a quantity able to quantify the uncertainty, or the randomness of a system. Information theory was initially developed by C. E. Shannon to quantify the expected value of the information contained in a message, usually in units such as bits \cite{shannon}. However, the Shannon entropy, used in standard top-down decision trees, does not guarantee the best generalization (see, for example, Ref.~\cite{maszczyk}). The Shannon entropy has then been generalized, successively. A. R{\'e}nyi introduced the most general definition of information measures that preserve the additivity for independent events and are compatible with the axioms of probability \cite{Renyi1}. C. Tsallis, on the contrary, introduced a nonadditive entropy, such as nonextensive statistical mechanics, generalizing then the Boltzmann-Gibbs theory \cite{Tsallis1}. R{\'e}nyi's and Tsallis's entropies are algebraically related, and both definitions include the Shannon entropy as a limit case. A wide spectrum of natural, artificial and social complex systems are now analyzed by means of these two entropies. At present, information theory finds applications in broad areas of science such as, in neurobiology \cite{rieke}, the evolution of molecular codes \cite{huelsenbeck}, model selection in ecology \cite{burnham}, thermal physics \cite{jaynes}, plagiarism detection \cite{bennet} or quantum information \cite{franchini}.

\noindent The purpose of this work is to introduce and study a class of \emph{anisotropic} generalization of R{\'e}nyi's entropy (RE). An obvious reason is that, in some problems of complex systems, the phase space may be highly anisotropic. Let us illustrate this concept by the following simple
example. We consider two small identical particles of mass $m$, immersed in
a fluid governed by the Stokes law, with friction coefficient ${\tilde{\gamma
}}$, and subject to central two body force $-\nabla V(r)$. In this case, it is easy to check that the
problem is equivalent to the one of a single particle, of reduced mass
$\mu=m/2$, immersed in a bath with friction coefficients $\gamma
={\tilde{\gamma}}/2$. After little calculations, we see that in this case the
statistical properties of the center of mass $R=(r_{1}+r_{2})/2$ are
completely decoupled by those of the relative motion $r\equiv r_{1}-r_{2}$. It
is also easily checked that the velocity of the center of mass \textit{i.e.},
${\dot{R}}=({\dot{r}}_{1}+{\dot{r}}_{2})/2$, obeys to the Ornstein Uhlenbeck
Fokker-Planck equation, whereas the statistical properties of the relative
motion are described by the following Kramer equation
\[
\frac{\partial\rho(\mathbf{r},\mathbf{v},t)}{\partial t}=\left[
-\frac{\partial}{\partial\mathbf{r}}\mathbf{v}+\frac{1}{\mu}\frac{\partial
}{\partial\mathbf{v}}\left(  \gamma\mathbf{v}+\nabla V(r)\right)
+\frac{\gamma K_{B}T}{\mu^{2}}\frac{\partial^{2}}{\partial v^{2}}\right]
\rho(\mathbf{r},\mathbf{v},t)
\]
\noindent with $K_B$ denoting Boltzmann's constant. Moreover, $T$ and $\bf v$ stand for temperature and $\dot{\bf r}$, respectively. The steady state solution is of the Boltzmann distribution type
\cite{risken}
\begin{equation}
\rho_{stat.}(r,v)=\rho_{0}\exp[-V(r)/(K_{B}T)]\exp[-\mu v^{2}/(2K_{B}T)]
\label{Equation0}%
\end{equation}
\noindent We are interested in diffusion of particles in the presence of a
logarithmically growing potential. This potential has attracted much interest
since it serves as a model of several physical systems. For example, charged
particles near a long and uniformly charged polymers, are subject to a
logarithmic potential. Other examples of systems having potentials showing a
logarithm and power-law potentials are: in condensation in polyelectrolyte
solutions \cite{manning}, nanoparticles with arbitrary two-dimensional force
field \cite{cohen}, and vortex dynamics in the two-dimensional model
\cite{bray}. For a logarithmically growing potential $V(x)\asymp V_{0}%
\log(x/a)$, for $x\gg a$ and $V_{0}>0$, we get an equilibrium distribution
with power-law tail $\rho_{stat.}(x)\sim x^{-V_{0}/(K_{B}T)}$
\cite{marksteiner}. Ref.~\cite{ryabov} reports on the study of an over damped
motion of a Brownian particle in the logarithmic-harmonic potential. In
Ref.~\cite{dechant} it can be found the study of the trajectories of a
Brownian particle moving in a confining asymptotically logarithmic potential,
obeying the over damped Langevin equation with potential
\begin{equation}
V(r)=g\log(1+r^{2}) \label{ex9}%
\end{equation}
\noindent where the parameter $g>0$ specifies the strength of the attractive
potential. Notice that this situation may be realized in experiments
\cite{kessler}-\cite{blickle}. In this case, the stationary solution reads
\begin{equation}
\rho_{stat.}(r,v)=\rho_{0}\frac{1}{(1+r^{2})^{g/(K_{B}T)}}\exp[-\mu
v^{2}/(2K_{B}T)] \label{ex10}%
\end{equation}
\noindent Eq.~(\ref{ex10}) clearly shows that the steady-state solution has a
short tail in velocity and long tail in distance $r$. We shall show that this
anisotropy of the PDF in the $\{\mathbf{r},\mathbf{v}\}$ phase-space can be
retrieved from the MaxEnt principle applied to the new class of entropy here
introduced, subject to natural scale-invariant restrictions (see below). This anisotropy of the phase space is manifest in magnetically
confined plasmas and, in general, in the case of integrable systems subject or
not to perturbations. Another example is provided by the evolution of
dynamical systems under the effect of noise where we consider the extended
phase space of the\ system plus source of noise. Also in this quite general
example, the extended phase space is anisotropic. Our main task is to propose
a generalization of the R\'{e}nyi entropy (GRE) that, still preserving the
additivity, is able to treat these anisotropic situations. We prove that the
GRE provides a new set of Liapunov functionals (more exactly H theorems ) for a large class of
Fokker-Planck equations describing particles transport in a random physical
environment. In this case the GRE is always monotonic. We encounter this
situation when we study the dynamics of charged particles in random electric
and magnetic fields \cite{Balescu1}.

\noindent Our generalization of RE results from the reinterpretation of the RE
and MaxEnt principles in the terms geometrical concepts in the functional
spaces of all PDF in a given phase space. Let us now proceed by introducing a
rigorous definition of the GRE and illustrating its main properties. The
definition of GRE starts from the reformulation of RE in the geometric term of
norm (or pseudo norm). Starting from the initial axiomatic definition in the
case of discrete probability field, R\'{e}nyi proved that, for a fixed value
of parameter $q$, appearing in his set of axioms, in particular the "axiom
$5^{\prime}$" (see Ref.~\cite{Renyi1}), the form $S_{R,q}(p_{i})=\frac{q}%
{1-q}\log\left(  {\displaystyle\sum\limits_{i=1}^{N}}p_{i}^{q}\right)  ^{1/q}$
$\ $is unique, up to a multiplicative constant. In general, \textit{i.e.}
including the continuum case, when the probability is defined in the terms of
some (possibly preferred invariant) measure $dm(\mathbf{x})$ with PDF $\rho(\mathbf{x)}$ in the space $\Omega$ , the previous
definition of RE can be extended, as $S_{R,q}[\rho]=\frac{q}{1-q}\log\left[
{\displaystyle\int\limits_{\Omega}}\left[  \rho(\mathbf{x})\right]
^{q}dm(\mathbf{x})\right]  ^{\frac{1}{q}}$. In the particular case when $\Omega$ is a set with $N$ elements and $m$ is the counting measure the original form is obtained, the permutation symmetry from the original Axiom 1 (see Ref.~\cite{Renyi1}) became the invariance under transformations that preserve the measure $m$. Observe that $S_{R,q}[\rho]$ for $q>1$ can geometrically
be reinterpreted in terms of the norm $\left\Vert \rho\right\Vert _{q}$
\begin{equation}
S_{R,q}[\rho]=\frac{q}{1-q}\log\left\Vert \rho\right\Vert _{q}\quad
;\quad\left\Vert \rho\right\Vert _{q}=\left[  {\displaystyle\int
\limits_{\Omega}}\left[  \rho(\mathbf{x})\right]  ^{q}dm(\mathbf{x})\right]
^{\frac{1}{q}};~q>1 \label{l1}%
\end{equation}
\noindent with the norm, $\left\Vert \rho\right\Vert _{q}$ defining the
distance in the standard Lebesgue $L_{q}(\Omega,dm)$ spaces \cite{Rudin},
\cite{ReedSimon}. However, in the case $0<q<1$ we are no longer able to
interpret $\left\Vert \rho\right\Vert _{q}$ in Eq.(\ref{l1}), as a distance.
This problem may easily be overcome by observing that the functional
\[
N_{q}[\rho]={\displaystyle\int\limits_{\Omega}}\left[  \rho\mathbf{(x}%
)\right]  ^{q}dm(\mathbf{x})
\]
\noindent may be interpreted as a distance \cite{Rudin}, \cite{LuschgiPages}
and \cite{SGW} so, for $0<q<1$, the definition of entropy reads
\[
S_{R,q}[\rho]=\frac{1}{1-q}\log N_{q}[\rho]
\]
with $N_{q}[\rho]~$playing the role of "distance", but in the more complicated
space $L_{q<1}(\Omega,dm)$ \cite{Rudin}. We remark that $N_{q}[\rho]$ and
spaces $L_{q<1}$ were used for studying the steady state distributions of
linear stochastic differential equations \cite{SGW} or of the stable
distributions with heavy tail \cite{LuschgiPages}. However, it should be
mentioned that, due to the complexity of the formalism, mathematicians prefer
to transfer $f(x)\in L_{q<1}$ in the standard $L_{1}$ space by
$f(x)\rightarrow\left\vert f(x)\right\vert ^{q}\in L_{1}$ \cite{Rudin}. The
physical counterpart of this transformation is the Tsallis averaging rule
\cite{Tsallis1}: if $\rho(x)$ is a PDF then for averaging it use $\left[
\rho(x)\right]  ^{q}/\left\langle \left[  \rho(x)\right]  ^{q}\right\rangle $,
not $\rho(x)$. \noindent Under a natural set of restrictions (consisting in
normalization, fixing the expectation values, positivity) on $\rho
\mathbf{(x}),$ the MaxEnt principle, applied to $S_{R,q}(\rho)$ with $q<1,$
\ generates distribution functions with heavy tail. The set of the PDFs
satisfying the restrictions (this set of PDFs will be denoted with
$\mathcal{K}$) is always convex. According to this interpretation,
geometrically we have that, for $0<q<1$, the MaxEnt PDF is the PDF
$\in\mathcal{K}$ corresponding to the maximal distance from the origin
$\rho\equiv0$, and reversely, for $q>1$, it is the PDF $\in\mathcal{K}$,
closest to the origin $\rho\equiv0$. Despite the corresponding equations for
the Lagrange multipliers could be quite complicate, from general arguments on
convex analysis we have that the convexity of the functional $\left\Vert
\rho\right\Vert _{q}$ with respect to the variable $\rho$ (or, similarly, the
concavity of the functional $N_{q}[\rho]$ with respect to $\rho$), ensures the
uniqueness of the solution of the MaxEnt problem \cite{ConvexOptimization}.
These properties, as well as the extensivity of RE, will be preserved in our
definition of GRE. However, the axiom 1 in the original work of R\'{e}nyi,
\textit{i.e.} the symmetry of RE \cite{Renyi1}, will not be preserved in GRE.
The complicated aspect of GRE is compensated by the fact that it is possible
to obtain more complex MaxEnt distribution function starting from simple,
natural restrictions, in contrast with the Shannon and R\'{e}nyi entropies
where the MaxEnt PDF come from a tautological transcription of the
restrictions. It is worth mentioning that, more complex generalizations of GRE
are possible, and the study in this respect is in progress \cite{sonnstei}.

\noindent\textbf{Definition of the GRE}.

\noindent In the case of anisotropic situations, it is quite natural to
introduce the new generalized entropy by starting from the definition of the
(anisotropic) norm of functions of many variables. To this end, we introduce
the norm of functions in the generalized $L_{p}$ spaces \cite{Besov}. Suppose
for simplicity that the variable components $\mathbf{x}$ from phase space
$\Omega$ can be split like $\mathbf{x}=\{x_{1},...,x_{p},x_{p+1},...,x_{n}\}$
or $\mathbf{x}=\{\mathbf{y,z}\}$ with $\mathbf{y=}\{x_{1},...,x_{p}\}$ and
$\mathbf{z=}\{x_{p+1},...,x_{n}\}$. The measure is also factorized $dm(\mathbf{x)}=dm(\mathbf{y,z})=dm_{y}(\mathbf{y})dm_{z}%
(\mathbf{z})$.

\noindent More general splitting, grouping, fits in this scheme but we restrict here to
the 2 subsets. In analogy with Eq.(\ref{l1}), we define for $p_{y},p_{z}>1$ the anisotropic norm as \cite{Besov}  
\begin{equation}
\left\Vert f\right\Vert _{p_{y},p_{z}}    =\left[  {\displaystyle\int
\limits_{\Omega_{y}}}dm_{y}(\mathbf{y)}\left[  \left[  {\displaystyle\int
\limits_{\Omega_{z}}}dm_{z}(\mathbf{z)}\left\vert f(\mathbf{y,z})\right\vert
^{p_{z}}\right]  ^{1/p_{z}}\right]  ^{p_{y}}\right]  ^{1/p_{y}}\label{l3}
\end{equation}
\noindent The functional $\left\Vert f\right\Vert _{p_{y},p_{z}}$ is
\emph{convex with respect to }$f$. The corresponding new entropy is defined as 
\begin{equation}
S_{p_{y},p_{z}}^{(1)}[\rho]   =\frac{p_{y}}{1-p_{z}}\log\left\Vert
\rho\right\Vert _{p_{y},p_{z}};~p_{y},p_{z}>1 \label{l4}
\end{equation}
\noindent Similarly, for $0\,<q_{y}<1$, $0<q_{z}<1$
we have a \emph{concave functional } $N_{q_{y},q_{z}}(f)$ that, in analogy of
the standard R\'{e}nyi case, may also be interpreted as the \textit{distance}
in the corresponding functional space and the corresponding entropy :
\begin{align}
N_{q_{y},q_{z}}(f)  &  ={\displaystyle\int\limits_{\Omega_{y}}}dm_{y}%
(\mathbf{y)}\left[  {\displaystyle\int\limits_{\Omega_{z}}}dm_{z}%
(\mathbf{z)}\left\vert f(\mathbf{y,z})\right\vert ^{q_{z}}\right]  ^{q_{y}%
}\label{l5}\\
S_{q_{y},q_{z}}^{(2)}[\rho]  &  =\frac{1}{1-q_{z}}\log N_{q_{y},q_{z}}%
(\rho);0\,<q_{y},q_{z}<1 \label{l6}%
\end{align}
\noindent The distance between PDF's $\rho_{1},\rho_{2}$ is $d(\rho_{1},\rho_{2}):=\left\Vert
\rho_{1}-\rho_{2}\right\Vert _{p_{y},p_{z}}$ \ for $p_{y},p_{z}>1$, and
$d(\rho_{1},\rho_{2}):=N_{q_{y},q_{z}}(\rho_{1}-\rho_{2})$ for $0\,<q_{y},q_{z}<1$. 
Notice that, as shown in Ref.~\cite{Besov},
for $p_{y},p_{z}>1$, the function $d(\rho_{1},\rho_{2})$ preserves the
triangle inequality\footnote{A similar method illustrated by Rudin in his
textbook \cite{Rudin} may be adopted for showing the validity of the triangle
inequality also for $0\,<q_{y},q_{z}<1$.}
\[
d(\rho_{1},\rho_{3})\leq d(\rho_{1},\rho_{2})+d(\rho_{2},\rho_{3})
\]
\noindent The norm $\left\Vert \rho\right\Vert _{p_{y},p_{z}}$ is a convex
functional whereas $N_{q_{y},q_{z}}[\rho]$ is concave
functional\footnote{i.e., for $0\leq\alpha\leq1$ we have for $\rho
(\mathbf{y,z)}=\alpha\rho_{1}(\mathbf{y,z)+}(1-\alpha)\rho_{2}(\mathbf{y,\ z)}%
$
\begin{align*}
\left\Vert \rho\right\Vert _{p_{y},p_{z}}  &  \leq\alpha\left\Vert \rho
_{1}\right\Vert _{p_{y},p_{z}}+(1-\alpha)\left\Vert \rho_{2}\right\Vert
_{p_{y},p_{z}}\\
N_{q_{y},q_{z}}[\rho]  &  \geq\alpha N_{q_{y},q_{z}}[\rho]+(1-\alpha
)N_{q_{y},q_{z}}[\rho]
\end{align*}
}. These properties give the intuitive geometric interpretation of MaxEnt
problem subject to linear constraints, in the framework of convex analysis
\cite{ConvexOptimization}. It follows that GRE is related to the geometry of
generalized Lebesgue space $L_{p_{y},p_{z}}$ consisting in the set of
functions $f(y,z)$ such that $N_{q_{y},q_{z}}[f]$ or $\left\Vert f\right\Vert
_{p_{y},p_{z}}$ are finite (like the R\'{e}nyi entropy in the Lebesgue space
$L_{p}$). Convexity properties imply the uniqueness of the solution of MaxEnt
problem with restrictions, despite the equations for Lagrange multipliers
could be very complex \cite{Comte}. Instead of working with two different
definitions of entropy (for two separate cases $0\,<q_{y},q_{z}<1$ and
$q_{y},q_{z}>1$) we prefer to compact the definitions in only one expression.
To this end, we observe that, for fixed $\rho\geq0$, both functions
$\left\Vert \rho\right\Vert _{p_{y},p_{z}}^{p_{y}}$ and $N_{q_{y},q_{z}}%
(\rho)~$are analytic in the variables $p_{y},p_{z}$ , $q_{y},q_{z}$ (at least
near the positive real axis) so we can do a unique analytic continuation in
their formula outside their initial domains in the following manner
\begin{align}
N_{p_{y}/p_{z},p_{z}}[\rho] &  =\left\Vert \rho\right\Vert _{p_{y},p_{z}%
}^{p_{y}}\label{l7}\\
S_{_{p_{y}/p_{z},p_{z}}}^{(2)}[\rho] &  =S_{p_{y},p_{z}}^{(1)}[\rho
]\label{l7.1}%
\end{align}
\noindent For compactness of the formulae, we use for all $q_{y},q_{z}>0$, and
$q_{z}\neq1$
\begin{align}
N_{q_{y},q_{z}}[\rho]  &  ={\displaystyle\int\limits_{\Omega_{y}}}%
dm_{y}(\mathbf{y)}\left[  {\displaystyle\int\limits_{\Omega_{z}}}%
dm_{z}(\mathbf{z)}\left\vert f(\mathbf{y,z})\right\vert ^{q_{z}}\right]
^{q_{y}}\label{l7.3}\\
S_{q_{y},q_{z}}^{(2)}[\rho]  &  =\frac{1}{1-q_{z}}\log N_{q_{y},q_{z}}(\rho)
\label{l7.4}%
\end{align}
\noindent Remark that the Axiom $1$, the symmetry, invariance under permutations for R{\'e}nyi entropy \cite{Renyi1} appears in a more general form: the invariance under transformations that acts independently in the spaces $\Omega_y$ and $\Omega_z$ that preserves the measures $m_y$ respectively $m_z$ . In the example given below, passive advection-diffusion of a tracer in turbulent field, the variables $y$ and measure $m_y$ are related to the statistical properties of a macroscopic, external, given turbulent velocity field, while $z$ and the measure $m_z$ give a statistical description of the effects of molecular diffusion. In this case is no symmetry transformation that mixes these very di¤er- ent type of variables, rather it is meaningful to relate this asymmetry of the GRE to the hierarchical relation between multiple scales, or causality effects, between spaces $\Omega_y$, $\Omega_z$.

\noindent\textbf{Properties of the GRE}

\noindent Notice that in the limit case $q_{y}\rightarrow1$ we obtain the
standard R\'{e}nyi entropy
\[
S_{1,q_{z}}^{(2)}[\rho]=\frac{1}{1-q_{z}}\log{\displaystyle\int\limits_{\Omega
_{y}}}dm_{y}(\mathbf{y}){\displaystyle\int\limits_{\Omega_{z}}}dm_{z}%
(\mathbf{z})\left\vert \rho(\mathbf{y,z})\right\vert ^{q_{z}}%
\]
and for $p_{z}\rightarrow1$ the Shannon entropy
\[
\underset{p_{z}\rightarrow1}{\lim}~\underset{p_{y}\rightarrow1}{\lim}%
S_{p_{y},p_{z}}^{(2)}\left\{  \rho\right\}  =-{\displaystyle\int
\limits_{\Omega_{y}}}dm_{y}(\mathbf{y)}{\displaystyle\int\limits_{\Omega_{z}}%
}dm_{z}(\mathbf{z)}\rho(\mathbf{y,z})\log\rho(\mathbf{y,z})
\]
\noindent We would like to underline that, if we perform the following scaling
of the variables $\mathbf{y}\rightarrow\alpha\mathbf{y}$, $\mathbf{z}%
\rightarrow\beta\mathbf{z}$, and the measures transform like as $dm_{y}%
(\mathbf{y})\rightarrow\alpha^{d_{1}}dm_{y}(\mathbf{y})$,$~dm_{z}%
(\mathbf{z})\rightarrow\beta^{d_{2}}\ dm_{z}(\mathbf{z})$, then the previously
defined entropies changes by constant. In this context, the variation of the
GRE is invariant under scaling, exactly like as in the case of the Shannon
entropy. In addition, notice that the GRE is extensive, like the R\'{e}nyi
entropy, because the norm $\left\Vert \rho\right\Vert _{p_{y},p_{z}}$ and the
functional $N_{q_{y},q_{z}}[\rho]$ are multiplicative, in analogy with
properties of the norm in the $L_{p}$ space. 

\noindent {\bf The MaxEnt principle}

\noindent The probability distribution functions may be obtained by the MaxEnt principle. Here, we shall determine the PDF by generalizing the calculations made for the case of the Shannon entropy, subject to the most general scale-invariant restrictions \cite{sonninoPRE}. To this end, we maximize the GRE, $S_{p_{y},p_{z}}^{(1,2)}[\rho]$, subject to the constraints
\begin{align}\label{pozitivity}
&{\displaystyle\int\limits_{\Omega_{y}}}dm_{y}(\mathbf{y)}{\displaystyle\int
\limits_{\Omega_{z}}}dm_{z}(\mathbf{z)}\rho(\mathbf{y,z})f_{k}(\mathbf{y,z})
  =c_{k};~1\leq k\leq M\\
&\rho(\mathbf{y,z}) \geq0\quad ; \quad f_{0}(\mathbf{y,z})  =1\ ;\ c_{0}=1
\end{align}
\noindent This means to find the extrema of $N_{q_{y},q_{z}}[\rho]$. From
Kuhn-Tucker theorem for maximization \cite{KuhnTucker}, we get
\begin{align}
&  \frac{\delta}{\delta\rho(\mathbf{y},\mathbf{z})}\left\{  N_{p_{y},p_{z}
}[\rho]+\!\!\!\!\!{\displaystyle\int\limits_{\Omega_{y}\times\Omega_{z}}}dm_{y}
(\mathbf{y)}dm_{z}(\mathbf{z)}\rho\mathbf{(\mathbf{y,z})}\left[
\mu(\mathbf{y,z})-{\displaystyle\sum\limits_{k=0}^{N}}\lambda_{k}
f_{k}(\mathbf{y,z})\right]  \right\}  =0\nonumber\\
&  \mu(\mathbf{y,z})\geq0;~\mu(\mathbf{y,z})\rho\mathbf{(\mathbf{y,z})=0}
\end{align}
\noindent where $\mu (y,z)$ and $\lambda_\kappa$ and are the multipliers corresponding to the positivity inequality and the linear restrictions, respectively. Here, we consider only the case $0<p_{y},p_{z}<1$. We introduce the notations
\begin{align*}
&  g(\mathbf{\lambda,y},\mathbf{z}):=\frac{1}{p_{y}p_{z}}{\displaystyle\sum
\limits_{k=0}^{N}}\lambda_{k}f_{k}(\mathbf{y,z})\\
&  a:=\frac{1-p_{y}}{1-p_{y}p_{z}}\quad;\quad b:=\frac{p_{z}}{1-p_{z}}%
\quad;\quad h(\mathbf{\lambda,y)}\mathbf{:}\mathbf{=}{\displaystyle\int
\limits_{\Omega_{z}}}dm_{z}(\mathbf{z}^{\prime}\mathbf{)}\left\vert
g(\mathbf{\lambda,y},\mathbf{z}^{\prime})\right\vert ^{-b}%
\end{align*}
\noindent By straightforward calculations, we get
\begin{equation}
\rho(\mathbf{\lambda,y},\mathbf{z})=g(\mathbf{\lambda,y},\mathbf{z}
)^{1/\left(  p_{z}-1\right)  }\left\vert h(\mathbf{\lambda,y)}\right\vert
^{-a} \label{f1}%
\end{equation}
\noindent Consider the particular case:
\begin{equation}
{\displaystyle\int\limits_{\mathbb{R}^{2}}}dydz\rho(y,z)y^{2}=c_{1}\quad
;\quad{\displaystyle\int\limits_{\mathbb{R}^{2}}}dydz\rho(y,z)z^{2} =c_{2}
\label{rest1}%
\end{equation}
\noindent From Eq.(\ref{f1}), we obtain (up to a multiplicative constant)
\begin{equation}
\rho(\lambda,y,z)=\frac{\left(  1+(\lambda_{1}y)^{2}\right)  ^{m}}{\left(
1+(\lambda_{1}y)^{2}+(\lambda_{2}z)^{2}\right)  ^{\frac{1}{1-p_{z}}}}
\quad;\quad m=-\frac{1-3p_{z}}{2(1-p_{z})}\frac{1-p_{y}}{1-p_{y}p_{z}} \label{rest2}%
\end{equation}
\noindent which corresponds to a PDF with different tail-exponents in the
variables $y,z$. If to Eqs.~(\ref{rest1}), we add the supplementary restriction
\begin{equation}
{\displaystyle\int\limits_{\mathbb{R}^{2}}}dydz\rho(y,z)y^{2}z^{2} =c_{3}
\label{rest3}
\end{equation}
\noindent it is possible to find a combination of the Lagrange multipliers
such that, up to a multiplicative constant, we get
\[
g(\mathbf{\lambda,y},\mathbf{z})=(1+a_{1}y^{2})(1+a_{2}z^{2})
\]
So that
$c_{1,2}->a_{1,2}$)
\begin{equation}
\rho(y,z)=K\frac{1}{\left(  1+a_{1}y^{2}\right)  ^{\kappa_{y}}}\frac
{1}{\left(  1+a_{2}z^{2}\right)  ^{\kappa_{z}}}\ ;\ \kappa_{y}=1/(1-p_{y}
p_{z})\ ;\ \kappa_{z}=1/(1-p_{z}) \label{greex1}
\end{equation}
\noindent  Hence, by putting $c_{2}=(1-p_{z})/\sigma^{2}$, and in the limit
case $p_{z}\rightarrow1$ (with the rest of the parameters kept constant), we
get
\begin{equation}\label{greex2}
\rho(y,z)=K\frac{1}{\left(  1+c_{1}y^{2}\right)  ^{\kappa_{y}}}\exp(-z^{2}
/\sigma^{2})
\end{equation}
\noindent which is exactly the form of the stationary solution analyzed in our
example Eq.(\ref{ex10}). Let us now consider the problem of variation of GRE
in a dynamical system, whose microscopic statistical features are described by
the Fokker Planck equation, when additional random effects, due to turbulence
at macroscopic scale, are taken into account by a random variable $\omega$. In
general, the diffusion term describes the effect of the interactions at the
atomic scale. The space of the additional random variable $\omega$, will be
denoted simply by $\Omega$. $\Omega$ corresponds to the previous $\Omega_{y}$,
but now it describes the effect of turbulent environment. The previous space
$\Omega_{z}$ is the usual phase space of the dynamical system, with
coordinates $\mathbf{z}=\{z_{1},...,z_{m}\}$. The typical example is the
passive advection-diffusion of tracer by a velocity field with turbulent
components and molecular diffusion. Consider the case when the evolution is
modeled by the more general advection-diffusion stochastic differential
equation (SDE) driven by the white noise $\zeta_{i}(t)$
\begin{align}
&  dz_{i}/dt=V_{i}(\mathbf{z},\omega)\ +\zeta_{i}(t)\quad ;\qquad 1\leq i\leq
m\label{fp9}\\
&  \left\langle \zeta_{i}(t_{1})\zeta_{j}(t_{2})\right\rangle =2D_{i,j}
(\mathbf{z},\omega)\delta(t_{1}-t_{2})\nonumber
\end{align}
\noindent where $V_{i}(\mathbf{z},\omega)$ and $D_{i,j}(\mathbf{z},\omega)$
satisfy the conditions
\begin{equation}
\label{fp11}\frac{\partial V_{i}(\mathbf{z},\omega)}{\partial z_{i}}
=0\quad;\quad\frac{\partial D_{i,j}(\mathbf{z},\omega)}{\partial z_{i}} =0
\end{equation}
\noindent where he convention of summation on repeated indices is adopted. The
corresponding Fokker-Planck equation, for a fixed $\omega$ is
\begin{equation}
\frac{\partial\rho(t,\omega,\mathbf{z})}{\partial t}=-\frac{\partial}{\partial
z_{i}}\left(  V_{i}\rho\right)  +\frac{\partial^{2}}{\partial z_{i}\partial
z_{j}}\left(  D_{i,j}\rho\right)  \label{fp13}
\end{equation}
\noindent Notice that the first condition of Eqs~(\ref{fp11}) is satisfied
\textit{e.g.}, by the must general Hamiltonian system, with $m/2$ degrees of
freedom (Liouville theorem). This general model contains some important
particular cases. When $\ m=3$ this corresponds to the passive tracer
transport by advection and molecular diffusion, in a turbulent flow, whose
statistical properties are encoded in the probability measure $dP(\omega)$. It
also describes the stochastic magnetic field line dynamics in tokamak
\cite{Balescu1}. For $m=2$ it may describe the transversal motion (transversal
to a constant magnetic field $\mathbf{B}$) of the charged particles in the
drift approximation, and subject to a random electric field $-\nabla
\phi(\mathbf{z},t,\omega)$ and collisions modelled by white noise $\zeta
_{i}(t)$ \cite{balescu3} .
\begin{align*}
&  dz_{i}/dt =\frac{1}{B}e_{i,j}\frac{\partial\phi}{\partial z_{j}}+\zeta
_{i}(t);~i=\overline{1,2}\\
&  \left\langle \zeta_{i}(t_{1})\zeta_{j}(t_{2})\right\rangle =\delta
_{i,j}\sigma\delta(t_{1}-t_{2})
\end{align*}
\noindent Here $e_{i,j}$ is the Levi-Civita symbol and $\sigma$ describe the effects of the collisions. 

\noindent We prove now the following important theorem, which is a sort of {\it H}-theorem describing the tendency of the GRE to increase in time. In general, we define the Liapunov function
$L(t):=\int\limits_{\Omega}dP(\omega)\left[  \int_{\Gamma}d^{m}\mathbf{x}
\left(  \rho(t,\omega,\mathbf{x})\right)  ^{p_{z}}\right]  ^{p_{y}}$ where
$\rho(t,\omega,\mathbf{z})$ is the solution of the Fokker-Planck equation
Eq.~(\ref{fp13}) for a fixed $\omega$. Then we have the following
\vskip0.1truecm
\noindent {\bf Proposition}

\noindent {\it Under the conditions Eqs~(\ref{fp11}), $\frac{d}{dt}L(t)(p_{z} -1)>0$ and the
corresponding GRE $S_{q_{y},q_{z}}^{(2)}[\rho]=\frac{1} {1-q_{z}}\log L(t)$ is
non decreasing in time}. 
\vskip0.1truecm

\noindent {\bf Proof :}
We start from the definition of the GRE and differentiate the expression with
respect to time. Then we use the Fokker-Planck equation for the time
derivative of $\rho$, and after integration by part in the $z$ coordinate, and by taking into account Eqs~(\ref{fp11}), we obtain
\begin{align*}
\frac{d}{dt}L(t) &  =\int\limits_{\Omega}dP(\omega)M(\omega,t)\int_{\Gamma
}d^{m}\mathbf{z}\left(  \rho(t,\omega,\mathbf{z})\right)  ^{p_{z}-2}
\frac{\partial\rho}{\partial z_{i}}\frac{\partial\rho}{\partial z_{j}}
D_{i,j}\\
M(\omega,t) &  =-p_{y}p_{z}(p_{z}-1)\left[  \int_{\Gamma
}d^{m}\mathbf{z}^{\prime}\left(  \rho(t,\omega,\mathbf{z}^{\prime})\right)
^{p_{z}}\right]  ^{p_{y}-1}
\end{align*}
with $\frac{\partial\rho}{\partial z_{i}}\frac{\partial\rho}{\partial z_{j}
}D_{i,j}\geq0$ from the second law of thermodynamics.
\vskip0.1truecm

\noindent In conclusion, we have introduced a generalization of the R{\'e}nyi entropy (GRE) that, still preserving the additivity, is able to treat dynamical systems in a highly anisotropic phase space. This is the case of magnetically confined plasmas or of integrable systems subject to perturbations. The anisotropy of the PDF in the phase-space can be retrieved from the MaxEnt principle applied to GRE, subject to natural scale-invariant restrictions. In these situations, the PDF may show different tail-exponents in the variables ; this property belongs only to GRE and not to the standard R{\'e}nyi entropy. We have also seen that the R{\'e}nyi and Shannon entropies are re-obtained by GRE as limit cases. Even though the extensivity of the RE is preserved in the GRE, the symmetry of the RE (axiom 1 of the RE), is not completely preserved in our generalized version. The symmetry group of R\'{e}nyi entropy, i.e. the measure preserving transformations of $\Omega_{y}\times\Omega_{z}$ splits into direct
product of measure preserving transformations of $\Omega_{y}$, respectively. This is the "price" we have to pay to treat dynamical systems in highly anisotropic phase-space. This point deserves attention. Strictly speaking, we can have the possibility to have a PDF with variables having different tails also in the Shannon theory (see the example illustrated at the beginning of the manuscript). However, in this case, it turns out that the PDF may be expressed in terms of a product of two PDFs, each of which depends on only one variable. The GRE may analyze examples "less trivial". We mean by this that the corresponding PDFs, obtained by the MaxEnt principle, may show different tail behaviours of the variables without necessarily being expressed in products of PDFs. However, as seen, the GRE includes the possibility to analyze simpler cases where the PDF is expressed as products of PDFs, each of which depends only on one variable, or on subgroups of independent variables [see Eq.~(\ref{greex2})].

\noindent The functionals that appears in the definition of the GRE may be interpreted as the distance in the corresponding functional space and in a wide range of the parameters $p_{y},p_{z}$ have useful concavity, respectively convexity, properties. Again, when the evolution of the system is modeled by the general advection-diffusion SDE driven by the white noise, we proved the validity of a sort of {\it H}-theorem, which results to be satisfied by the GRE when the velocity flows and the diffusion coefficients are {\it divergenceless} in the phase-space of the dynamical system [see Eq.~(\ref{greex2})].

\noindent This work gives several perspectives. Through the thermodynamical field theory (TFT) \cite{sonninoTFT}, it is possible to estimate the PDF when the nonlinear contributions cannot be neglected \cite{sonninopdf}. The next task should be to establish the relation between the reference, stationary PDF, derived by the MaxEnt principle applied to GRE, subject to scale-invariant restrictions, with the ones found by the TFT. 
\vskip0.2truecm

\noindent G. Sonnino is very grateful to M. Malek Mansour, of the Universit{\'e} Libre de Bruxelles, for his scientific suggestions and for his help in the development of this work. G. Steinbrecher acknowledges J. Misquich for useful discussions. We also acknowledge M. Van Schoor of the Royal Military School.

\end{document}